\documentclass[11pt]{book}

\usepackage{amssymb}
\usepackage{amsmath}
\usepackage{amsfonts}
\usepackage{epsfig}

\makeatletter

\renewenvironment{thebibliography}[1]
     {\section*{\bibname}%
      \@mkboth{\MakeUppercase\bibname}{\MakeUppercase\bibname}%
      \list{\@biblabel{\@arabic\c@enumiv}}%
           {\settowidth\labelwidth{\@biblabel{#1}}%
            \leftmargin\labelwidth
            \advance\leftmargin\labelsep
            \@openbib@code
            \usecounter{enumiv}%
            \let\p@enumiv\@empty
            \renewcommand\theenumiv{\@arabic\c@enumiv}}%
      \sloppy
      \clubpenalty4000
      \@clubpenalty \clubpenalty
      \widowpenalty4000%
      \sfcode`\.\@m}
     {\def\@noitemerr
       {\@latex@warning{Empty `thebibliography� environment}}%
      \endlist}

\makeatother

\newcommand{\sect}[1]{\setcounter{equation}{0}\section{#1}}

\renewcommand\bibname{References}

\begin{document}

\setlength{\baselineskip}{5.0mm}

%\tableofcontents

\chapter[Heavy-tailed random matrices]{Heavy-tailed random matrices}
\thispagestyle{empty}
 
\ \\

\noindent
{{\sc Z. Burda} and {\sc J. Jurkiewicz}  
\\~\\ Marian Smoluchowski Institute of Physics and \newline
Mark Kac Complex Systems Research Center, \newline
Jagiellonian University, \newline 
Reymonta 4, 30-059 Krak\'ow, Poland}

\begin{center}
{\bf Abstract}
\end{center}
We discuss non-Gaussian random matrices whose elements are random variables 
with heavy-tailed probability distributions. In probability theory heavy tails of the 
distributions describe rare but violent events which usually have dominant influence 
on the statistics. They also completely change universal properties of eigenvalues and 
eigenvectors of random matrices. We concentrate here on the universal macroscopic 
properties of (1) Wigner matrices belonging to the L\'evy basin of attraction, 
(2) matrices representing stable free random variables and (3) a class of heavy-tailed 
matrices obtained by parametric deformations of standard ensembles.

%%%%%%%%%%%%%%%%%%%%%%%%%%%%%%%%%%%%%%%%%%%%%%%%%%%%%%%%%%%%%%%%%%%%%%%%%%%
\sect{Introduction}\label{intro}

Gaussian random matrices have been studied over many decades and are well known by now 
\cite{Mehta}. Much less is known about matrices whose elements display strong fluctuations 
described by probability distributions with heavy tails.

Probably the simplest example of a matrix from this class is a real symmetric 
($A_{ij}\!=\!A_{ji}$) random matrix $A_N$ with elements $A_{ij}$, $1\le
i\le j\le N$, being 
independent identically distributed (i.i.d.) centered real random variables with a probability density function (p.d.f.) falling off as a power
\begin{equation}
p(x) \sim |x|^{-\alpha-1} \ .   
\label{alpha_pdf}
\end{equation} 
for $|x| \rightarrow \infty$.  
The smaller is the value $\alpha$ the heavier is the tail. Higher moments of this distribution do not exist.
For $\alpha \in (0,1]$ the tail is extremely heavy and the mean-value 
does not exist since the corresponding integral $\int x p(x) dx$ is divergent. For $\alpha \in (1,2] $ the mean-value does exist but the variance does not. The influence of heavy tails on the statistical properties of a random matrix is enormous. It is particularly apparent for $\alpha<1$. In this case the elements $A_{ij}$ assume values scattered over a wide range which itself quickly increases when $N$ goes to infinity. The largest element $|a_{max}|$ of the matrix is of the order $|a_{max}| \sim N^{2/\alpha}$ and its value strongly fluctuates from matrix to matrix. The distribution of the normalized value of 
the maximal element $x=|a_{max}|/N^{2/\alpha}$ is given by a Fr\'echet distribution which itself has a heavy-tail. The largest element of the matrix may be larger than the sum of all remaining ones. The values of the elements change so dramatically from matrix to matrix that one cannot speak about a typical matrix or about self-averaging for large $N$. In the limit $N \rightarrow \infty$ matrices $A_N$ may look effectively very sparse\footnote{
Sparse random matrices are discussed in Chapter 23 of this book.}. 
Indeed if one considers a rescaled matrix $A_N/|a_{max}|$ one will find that only a finite fraction of all elements of this matrix will be significantly different from zero. This effective sparseness is quantified by the inverse participation ratio $Y_2$ constructed from normalized weights $w_{ij} = |a_{ij}|/\sum_{ij} |a_{ij}|$, which sum up to unity $\sum_{ij} w_{ij} = 1$,
\begin{equation}
Y_2 = \frac{2}{N(N+1)} \overline{\sum_{i\le j} w^2_{ij}} \ .
\end{equation}
The bar denotes the average 
over matrices $A_N$. In the limit $N\rightarrow \infty$ the participation ratio is  $Y_2 = 1 - \alpha > 0$  for $\alpha \in (0,1)$ \cite{BM}. This means that only a finite fraction of matrix elements is relevant in a given realization of the matrix. This is a completely different behavior than the one known from considerations of Gaussian random matrices. For $\alpha \le [1,2)$, although
$Y_2=0$ in the limit $N\rightarrow \infty$, one still observes very
large fluctuations of individual matrix elements which in particular
lead to a localization of eigenvectors that will be shortly discussed
towards the end of the next section. Only for $\alpha > 2$ the behavior
of matrices resembles\footnote{The convergence to the limiting
semicircle law is generically very slow in the presence of power-law
tails and moreover for $\alpha\le 4$ microscopic properties are
significantly different than for generic Gaussian matrices as we will
shortly mention later.} that known for Gaussian matrices. In this case,
the variance $\sigma^2$ of (\ref{alpha_pdf}) is finite and the
eigenvalue density of the matrix $A_N/\sqrt{N}$ converges for
$N\rightarrow \infty$ to the Wigner semicircle law $\rho(\lambda) =
\sqrt{4\sigma^2 - \lambda^2}/(2\pi\sigma^2)$, independently of the details of the probability distribution. Random matrices having the same limiting eigenvalue density for $N\rightarrow \infty$ are said to belong to the same macroscopic universality class. For $\alpha>2$ it is called a Gaussian universality. This class is very broad and comprises a whole variety of random matrices. In particular one can prove \cite{Pastur} that if $A_N$ is a symmetric random matrix with independent (but not necessarily identically distributed) centered entries with the same variance $\sigma^2$, then the condition for the eigenvalue distribution of $A_N/\sqrt{N}$ to converge to the Wigner semicircle law is
\begin{equation}
\lim_{N\rightarrow \infty} 
\frac{1}{N^2} \sum_{i\le j} \int_{|x| > \epsilon \sqrt{N}} x^2 p_{ij}(x) \, dx  = 0
\end{equation}
where $\epsilon$ is any positive number and $p_{ij}(x)$ is the p.d.f. for the $ij$-th element of the matrix.
As a matter of fact this condition is almost identical as the Lindeberg condition
known from the central limit theorem for the distribution of a sum of random numbers to converge to a normal distribution \cite{Feller}. The fastest convergence to the limiting semicircle law is achieved for matrices whose elements are independent Gaussian random variables. A prominent place in this macroscopic universality class is taken by the ensemble of symmetric Gaussian matrices whose diagonal elements have a twice bigger variance than the off-diagonal ones ${\cal N}(0,\sigma^2(1+\delta_{ij}))$. Clearly, such matrices fulfill 
the Linderberg condition. The probability measure in the ensemble of such matrices can be written as 
\begin{equation}
d\mu(A) = D A \exp - \frac{1}{2\sigma^2} {\rm tr} A^2 
\end{equation}
where $D A$ is a flat measure $D A = \prod_{1\le i\le j\le N} d A_{ij}$. 
The measure $d\mu(A)$ is manifestly invariant with respect to the
orthogonal transformations: $A \rightarrow O A O^T$, where $O$ is an
orthogonal matrix. This is the GOE ensemble. In the limit $N \rightarrow
\infty$ the eigenvalue density of $A_N/\sqrt{N}$ approaches a well-known semicircle distribution. In a similar way one can construct GUE and GSE ensembles which are extensively discussed in other chapters of the book.  In this chapter we will be mostly interested in matrices for which the variance does not exist.

\sect{Wigner-L\'evy matrices}

In this section we will discuss properties of heavy tailed symmetric matrices with i.i.d. elements (\ref{alpha_pdf}) for $\alpha \in (0,2)$.
We call them Wigner-L\'evy matrices since the L\'evy distribution is the corresponding stable law for $0<\alpha < 2$ which plays an analogous role from the point of view of the central limit theorem as the Gaussian distribution for $\alpha \ge 2$ \cite{GK}. The L\'evy distributions are sometimes called $\alpha$-stable laws.

Before we discuss Wigner-L\'evy matrices let us briefly recall  basic
facts about L\'evy distributions. In addition to a stability index $\alpha$ these laws are characterized by an asymmetry parameter $\beta \in [-1,1]$, which will be discussed below, and a scale parameter $R>0$, called the range, which plays a similar role as the standard deviation $\sigma$ for the normal law.
The statement that a L\'evy distribution is a stable law (with respect to addition) means that a sum of two independent L\'evy random variables $x=x_1+x_2$ with a given index $\alpha$ is again a L\'evy random variable with the same $\alpha$. The range $R$ and asymmetry $\beta$
of the resulting distribution can be calculated as
\begin{equation}
R^\alpha = R_1^\alpha + R_2^\alpha,\quad \beta=\frac{\beta_1 R_1^\alpha + \beta_2 R_2^\alpha}{R_1^\alpha + R_2^\alpha} \ .
\label{conv} 
\end{equation}
For $\beta_1=\beta_2$ the asymmetry is preserved and the relation for the effective range is a generalization of the corresponding one for independent Gaussian random variables, where the sum is also a Gaussian random variable with the variance $\sigma^2 = \sigma^2_1+\sigma^2_2$. Actually for $\alpha=2$ the range and the standard deviation are related  
as $\sigma = \sqrt{2} R$. 

The p.d.f. $L^{R,\beta}_\alpha(x)$ of the L\'evy distribution with the stability
index $\alpha$, the asymmetry $\beta$ and the range $R$ is conventionally 
written as a Fourier transform of the characteristic function, since its form is known explicitly. There are several definitions used in the literature, here we quote one, which seems to be the most common \cite{GK,Nolan}
\begin{equation}
L^{R,\beta}_\alpha(x) = \frac{1}{2\pi} \int_{-\infty}^{+\infty} d k 
\widehat{L}^{R,\beta}_\alpha(k)  e^{-ikx}
\label{FT}
\end{equation}
where\footnote{For $\alpha=1$ it assumes a slightly form: $c(k) =
-R|k|\left( 1+ i (2\beta/\pi) {\rm sgn}(k) \ln (Rk)\right)$.}
\begin{equation}
c(k) = \ln \widehat{L}^{R,\beta}_\alpha(k) = -R^\alpha |k|^\alpha \left(
1 + i \beta {\rm sgn}(k) \tan (\pi\alpha/2)\right) \ .
\label{charL}
\end{equation}
The logarithm of the characteristic function $c(k)$ is usually called a cumulant-generating function. This name is slightly misleading since for L\'evy distributions only the first cumulant exists for $\alpha \in (1,2)$ or none for $\alpha \in (0,1]$. Therefore we will rather call it the $c$-transform. The characteristic function is known explicitly in contrast to the corresponding p.d.f. $L^{R,\beta}_\alpha(x)$ which can be expressed in terms of simple functions only for $\alpha=1$, $\beta=0$ (Cauchy distribution) $\alpha=3/2$, $\beta=\pm 1$ (Smirnoff distribution). For $\alpha=2$, the characteristic function (\ref{charL}) becomes a Gaussian function independent of $\beta$. A stability of the L\'evy distribution can be easily verified. A p.d.f. for the sum of two independent variables $x_{1+2}=x_1+x_2$ is a convolution of the p.d.f.'s for individual components $x_1$ and $x_2$ and thus the corresponding characteristic function is a product of two characteristic functions. It is easy to check by inspection of (\ref{charL}) that the relations (\ref{conv}) are indeed satisfied. It is less trivial to demonstrate that the Fourier transform of the characteristic function (\ref{FT}) is a non-negative function. Actually this is only the case for $0<\alpha \le 2$. L\'evy distributions $L^{R,\beta}_\alpha(x)$ for $\alpha \in (0,2]$ are the only stable laws. The asymmetry parameter $\beta$ controls the skewness of the distribution. For $\beta=0$, the characteristic function (\ref{charL}) is even and so is the p.d.f. $L^{R,0}_\alpha(x) = L^{R,0}_\alpha(-x)$. For $\beta \ne 0$ the p.d.f. is skew and has a different left and right asymptotic behavior 
\begin{equation}
L^{R,\beta}_\alpha(x) \stackrel{x\rightarrow \pm \infty}{\longrightarrow} 
(1 \pm \beta ) \frac{\gamma_\alpha R^\alpha}{|x|^{\alpha+1}}
\end{equation}
where $\gamma_\alpha = \Gamma(1+\alpha) \sin(\pi \alpha/2) /\pi $.
In the extreme cases $\beta = \pm 1$ one of the tails is suppressed and for $\alpha< 1$ the distribution becomes fully asymmetric with a support only on the positive (resp. negative) semiaxis. 
One should note that the 
mean value of the L\'evy distribution for $\alpha>1$ equals zero, independently of the asymmetry $\beta$. For $\alpha=2$ the dependence on $\beta$ disappears. If one sets $R=1$ in (\ref{charL}) one obtains a standardized L\'evy distribution. A L\'evy random variable $x$ with an arbitrary $R$ can be obtained from the corresponding standardized one $x_*$ by a rescaling $x = Rx_*$, hence $L^{R,\beta}_\alpha(x) = L^{1,\beta}_\alpha(x/R)/R$. 

Let us now consider a symmetric matrix $A_N$ with elements $A_{ij}$ for
$1 \le i \le j \le N$ being i.i.d.  L\'evy random variables with the
p.d.f. $L^{R,\beta}_\alpha(x)$. We are now interested in the eigenvalue
density of such a matrix in the limit $N\rightarrow \infty$. As we shall
see below the eigenvalue density of the matrix $A_N/N^{1/\alpha}$
converges to a limiting density $\rho(\lambda)$ which is completely
different from the Wigner semicircle law and has an infinite support and
heavy tails. The choice of the scaling factor $N^{1/\alpha}$ is related
to the universal scaling properties of the L\'evy distribution and for
this reason can be viewed as a generalization of the scaling for
Gaussian random matrices ($\alpha\!=\!2$). The problem of the
determination of the limiting eigenvalue distribution of Wigner-L\'evy
matrices is not simple because the standard methods used for Gaussian
matrices or matrices from invariant ensembles do not apply here. A
special method tailored  to the specific universal features of
heavy-tailed distributions was necessary to attack this problem. Such a
method, being a beautiful adaptation of the cavity method which
maximally exploits universal properties of $\alpha$-stable
distributions, was invented in \cite{CB}. The description of the method
is beyond the scope of this chapter and we refer the interested reader
to the original paper \cite{CB} and to the paper \cite{BJNPZ} where the
derivation was explained step by step and some details were corrected.
Here we only quote the final result. The eigenvalue density $\rho(\lambda)$ of a Wigner-L\'evy matrix is given by a L\'evy function with an index $\alpha/2$, being a half of the index $\alpha$ of the p.d.f. $L^{R,\beta}_\alpha(x)$ used to generate the matrix elements, and an effective 
``running'' range $\widehat{R}(\lambda)$ and asymmetry parameter $\widehat{\beta}(\lambda)$
\begin{equation}
\rho(\lambda) = \frac{1}{\Lambda} 
\widehat{\rho}\left( \frac{\lambda}{\Lambda} \right) \quad {\rm where} \qquad
\widehat{\rho}(\lambda) = L^{\widehat{R}(\lambda),\widehat{\beta}(\lambda)}_{\alpha/2}(\lambda) 
\label{rho}
\end{equation} 
and 
\begin{equation}
\Lambda= R \left(\frac{\Gamma(1+\alpha) \cos(\pi \alpha/4)}{\Gamma(1+\alpha/2)}\right)^{1/\alpha}  \ .
\end{equation}
The scale parameter $\Lambda$ is proportional to the original range $R$.
The functions $\widehat{R}(\lambda)$, $\widehat{\beta}(\lambda)$ 
satisfy a set of integral equations
\begin{equation}
\widehat{R}^{\frac{\alpha}2}(\lambda) = 
\int_{-\infty}^{+\infty} dx \ |x|^{-\frac{\alpha}{2}} 
L^{\widehat{R}(x),\widehat{\beta}(x)}_{\alpha/2}(\lambda - x) 
\label{wr}
\end{equation}
\begin{equation}
\widehat{\beta}(\lambda) = 
\frac{\int_{-\infty}^{+\infty} dx \ {\rm sign}(x) |x|^{-\frac{\alpha}{2}} 
L^{\widehat{R}(x),\widehat{\beta}(x)}_{\alpha/2}(\lambda -x)}
{\int_{-\infty}^{+\infty} dx \ |x|^{-\frac{\alpha}{2}}
L^{\widehat{R}(x),\widehat{\beta}(x)}_{\alpha/2}(\lambda -x)}
\label{wb}
\end{equation}
where the integrals should be interpreted as Cauchy principal values.
The equation (\ref{wb}), which was derived in \cite{BJNPZ}, is a corrected  version of this equation given in \cite{CB}. 

We conclude the section with some comments. 

The dependence on $\lambda$ on the right hand side of (\ref{rho})
appears not only through the main argument of the function but also 
through the dependence of the effective parameters 
$\widehat{R}(\lambda)$ and $\widehat{\beta}(\lambda)$ on $\lambda$. 
This makes the resulting expression very complex. The equations 
for $\widehat{\beta}(\lambda)$ and $\widehat{R}(\lambda)$ cannot 
be solved analytically. It is also not easy to  
solve them numerically because the computation of the function
$L^{R,\beta}_\alpha(x)$, which is the main building block of the above
integral equations, 
is numerically unstable if one does the Fourier integral (\ref{FT}) in a straightforward way since its integrand is a strongly oscillating function. It is necessary to apply a non-trivial transformation of the integration contour in the complex plane to assure that the integration becomes numerically stable \cite{Nolan}. We skip the details here and again refer the interested reader to \cite{BJNPZ} for details. 

The eigenvalue density (\ref{rho}) holds for any $\alpha \in (1,2)$
and is independent of the asymmetry $\beta$ of the p.d.f. for matrix elements. The independence of $\beta$ might 
be surprising but it can be easily checked by inspection. Indeed, equations 
(\ref{wr}) and (\ref{wb}) depend only on $\alpha$ and do not involve any
dependence on the parameters $\beta$ or $R$ of the p.d.f. for matrix elements,  $L^{R,\beta}_\alpha(x)$ . Numerically the solution (\ref{rho}) seems to extend also to the range of $\alpha \in (0,1]$ but for symmetric distributions ($\beta=0$) only.

\begin{figure}
\begin{center}
\epsfig{figure=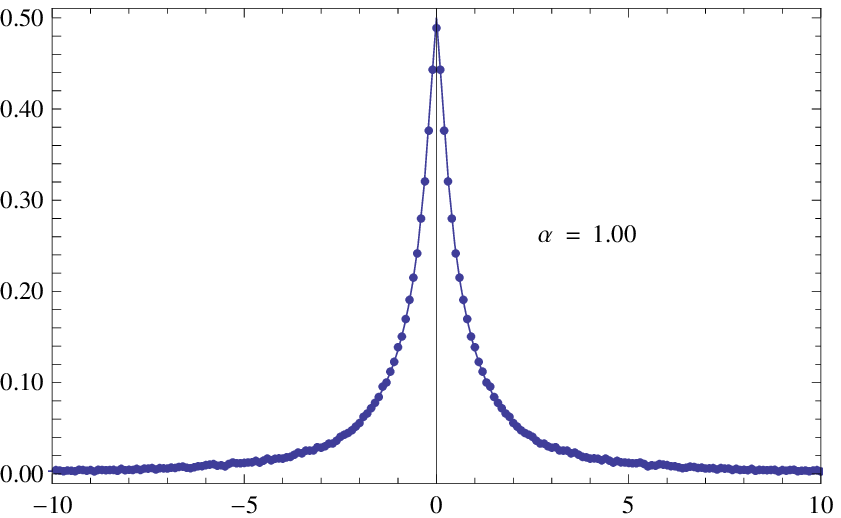 ,width=6.1cm} \  
\epsfig{figure=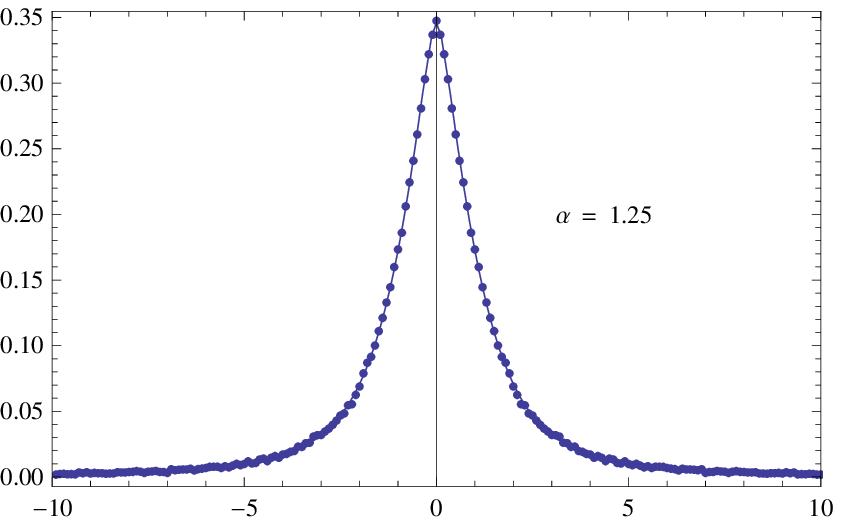 ,width=6.1cm}
\epsfig{figure=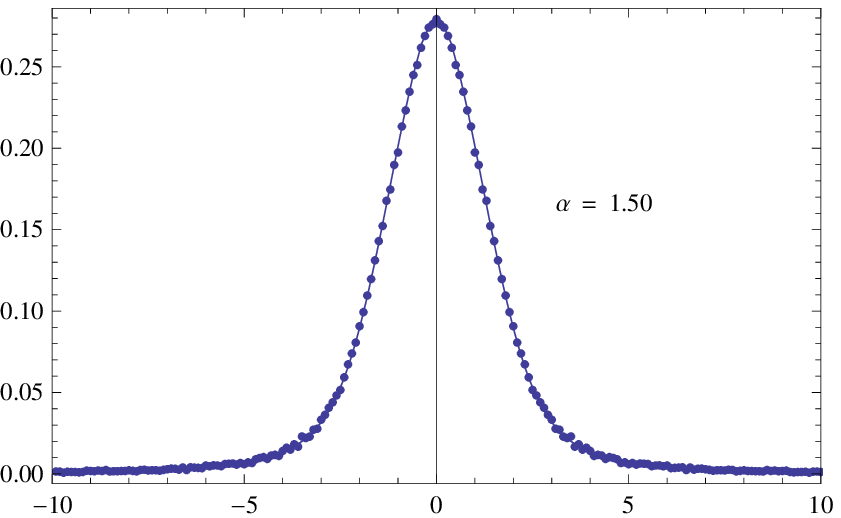 ,width=6.1cm} \ 
\epsfig{figure=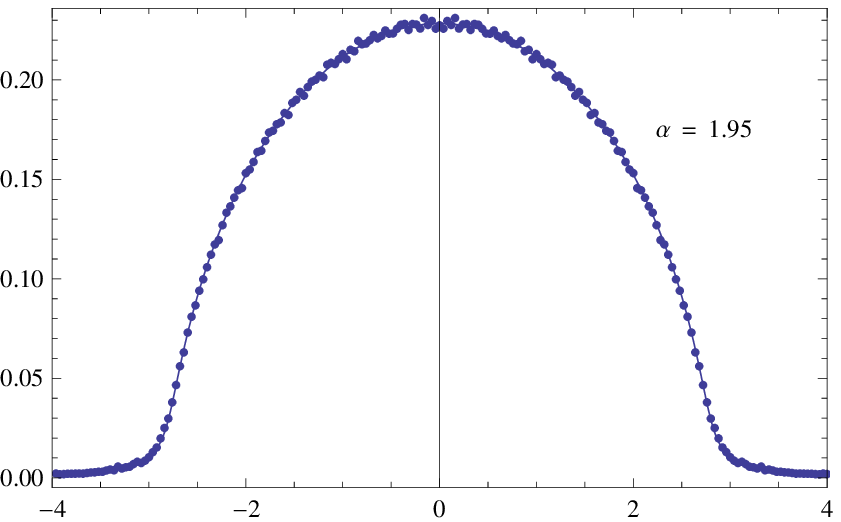 ,width=6.1cm}
\end{center}
\caption{\label{rho_fig} 
The eigenvalue density for infinite Wigner-L\'evy matrices with the index $\alpha=1.0,1.25,1.5,1.95$ and the range $R=1$ (solid line) and the corresponding numerical densities obtained by Monte-Carlo generated matrices of size $200\times 200$ with entries distributed according to $L^{1,0}_\alpha(x)$.}
\end{figure}

The limiting eigenvalue density $\rho(\lambda)$ is an unimodal even function.
The height of the maximum at $\lambda=0$ is 
\begin{equation}
\rho(0) = \frac{\Gamma(1+2/\alpha)}{\pi R} 
\left(\frac{\Gamma^2(1+\alpha/2)}{\Gamma(1+\alpha)}\right)^{1/\alpha} \ .
\end{equation}
For large $|\lambda| \rightarrow \infty$ the eigenvalue density has heavy tails
\begin{equation}
\rho(\lambda) \sim \frac{1}{\pi} \Gamma(1+\alpha) \sin\left(\frac{\alpha \pi}{2}\right) \frac{R^\alpha}{|\lambda|^{\alpha+1}}
\label{large_lambda_WL}
\end{equation}
with the same power as the p.d.f. for the matrix elements, although at first sight one might have the impression that the power should rather be $\alpha/2$ for it is the index of the L\'evy function on the right hand side of (\ref{rho}). This is not the case because also the dependence of the running parameters $\widehat{R}(\lambda)$ and $\widehat{\beta}(\lambda)$ on $\lambda$ contributes to a net asymptotic behavior.  
In Figure \ref{rho_fig} we show limiting distributions $\rho(\lambda)$ of the Wigher-L\'evy matrices with $R=1$ and different values of $\alpha$ and compare them to  eigenvalue histograms obtained numerically by Monte-Carlo generation of symmetric $200 \times 200$ matrices with i.i.d. entries with the p.d.f. $L^{1,0}_\alpha(x)$. The agreement is very good and finite size effects are small.

Another feature of Wigner-L\'evy matrices which distinguishes them 
from Wigner matrices from the Gaussian universality class is a localization of  the eigenvectors \cite{CB}. The degree of localization is measured by the inverse participation ratio 
\begin{equation}
y_2 = \sum_{i=1}^N \psi_i^4
\end{equation}
calculated for the elements $\psi_i$'s of a normalized eigenvector $\sum_{i=1}^N \psi_i^2 = 1$. For Wigner matrices from the Gaussian universality class $y_2=0$ in the limit $N\rightarrow \infty$ and it is independent of the corresponding eigenvalue. For Wigner-L\'evy matrices with the index $\alpha \in (1,2)$, the average inverse participation ratio 
$y_2=y_2(\lambda)$ depends on the eigenvalue $\lambda$.
For $\lambda$ smaller than a certain critical value, $\lambda\le \lambda_{cr}$, 
$y_2(\lambda)=0$, however for $\lambda>\lambda_{cr}$ above this value $y_2(\lambda)$ is positive and it grows monotonically to one when $\lambda$ increases \cite{CB}. This means that only a finite number of elements $\psi_i$ are significantly different from zero or phrasing it differently that the vector is localized on a subset of cardinality of the order $1/y_2(\lambda)$.  
When $\lambda$ goes to infinity $y_2(\lambda)$ tends to one. This means that for extremely large eigenvalues all but one elements of the corresponding vector are equal zero and the eigenvector is localized on exactly one state. The critical value $\lambda_{cr}$ depends on $\alpha$ and grows monotonically from zero for $\alpha=1$ to infinity for $\alpha=2$.  The fact that $\lambda_{cr}=0$ for $\alpha=1$ means that in
this case (and also for $\alpha<1$) all eigenvectors are localized.

The eigenvalue density (\ref{rho}) defines a whole universality class of
Wigner-L\'evy matrices. It is a counterpart of the Wigner semicircle law. 
The eigenvalue density of a matrix $A_N/N^{1/\alpha}$ 
with i.i.d. entries with a p.d.f. $p(x)$ will converge to the same limiting law if the function $p(x)$ has the same asymptotic behavior as $L^{R,\beta}_\alpha(x)$. More precisely, if the p.d.f. $p(x)$ is centered 
($\int_{-\infty}^{+\infty} dx x p(x) = 0$) for $\alpha \in (1,2)$
(or even for $\alpha \in (0,1]$) and has the following asymptotic behavior
\begin{equation}
p(x) \stackrel{x \rightarrow \pm \infty}{\longrightarrow} \frac{C_\pm}{|x|^{1+\alpha}}  
\end{equation}
then the eigenvalue distribution of the matrix $A_N/N^{1/\alpha}$ converges to 
the same limiting density $\rho(x)$ (\ref{rho}) as for the corresponding matrix with the p.d.f. $L^{R,\beta}_\alpha(x)$ with the same index $\alpha$ and
\begin{equation}
R = \left(\frac{C_+ + C_-}{2 \gamma_\alpha}\right)^{1/\alpha} \ , \quad 
\beta= \frac{C_+ - C_-}{C_+ + C_-} \ ,
\end{equation}
where as before $\gamma_\alpha = \Gamma(1+\alpha) \sin(\pi \alpha/2)/\pi$. 
One can probably extend this macroscopic universality to a class of matrices with 
independent but not necessarily identically distributed entries however 
having distributions with the same asymptotic behavior. 

We finish this section with a comment on microscopic properties of
Wigner matrices with heavy tails (\ref{alpha_pdf}). We have seen so far
that macroscopic properties of Wigner matrices change at $\alpha=2$. For
$\alpha>2$ Wigner matrices belong to the Gaussian universality class and
their eigenvalue density converges for large $N$ to the Wigner
semicircle law while for $\alpha<2$ they belong to the L\'evy
universality class and their eigenvalue density converges to the
limiting distribution given by (\ref{rho}). One may ask if $\alpha=2$ is
also a critical value for microscopic properties like for instance
eigenvalue correlations or the statistics of the largest eigenvalue
$\lambda_{max}$. This question has already been partially studied. It
was found that in this case the critical value of the exponent $\alpha$
is rather $\alpha=4$ \cite{BBP1}. For $\alpha>4$ the largest eigenvalue
of the matrix $A_N/\sqrt{N}$, for $N\rightarrow \infty$, fluctuates
around the upper edge of the support of the Wigner semicircle
distribution and the fluctuations are of the order $N^{-2/3}$. A
rescaled quantity $x=(\lambda_{max}-\lambda_{edge}) N^{2/3}$ obeys the
Tracy-Widom statistics \cite{TW1,TW2} although the convergence to this
limiting distribution is rather slow. For $\alpha < 4$ the largest eigenvalue is of the order $N^{(4-\alpha)/(2\alpha)}$ and a rescaled quantity  $y=\lambda_{max}/N^{(4-\alpha)/(2\alpha)}$ is distributed according to a modified Fr\'echet law. For $\alpha=4$ which is a marginal case the two regimes are mixed in the proportions depending on details of the p.d.f. for matrix elements, in particular on the amplitude of the tail. Roughly speaking for $\alpha>4$ the eigenvalue repulsion shapes the microscopic properties of the matrix and leads to the Tracy-Widom statistics while for $\alpha<4$ the repulsion plays a secondary role. The dominant effect is in this case related to
fluctuations  in the tail of the distribution which are so large that the repulsion can be neglected and the largest eigenvalues can be treated as independent of each other. Actually it has been known for some time \cite{Sosh} that indeed the largest eigenvalues are given by a Poisson point process with a Fr\'echet intensity related to the statistics of the largest elements in the random matrix coming from the tail of the distribution (\ref{alpha_pdf}) for $\alpha\le 2$. In paper \cite{BBP1} an argument was given that basically the same picture holds also for $2< \alpha<4$. 

Wigner matrices are discussed in Chapter 21. The interested reader can find there a discussion also of other aspects of this class of random matrices.

\sect{Free random variables and free L\'evy matrices}

It is sometimes convenient to think of whole matrices as entities 
and to formulate for them probabilistic laws. One can ask for example
if one can calculate the eigenvalue density $\rho_{1+2}(\lambda)$
of a sum of two symmetric (or hermitian) $N\times N$ random matrices
\begin{equation}
A_{1+2} = A_1 + A_2
\label{M12}
\end{equation}
given the densities $\rho_1(\lambda)$ and $\rho_2(\lambda)$ of
$A_1$ and $A_2$. In general the answer to this question is 
negative since the resulting distribution depends on many other factors. 
The situation becomes less hopeless for large matrices. It turns out that 
for $N\rightarrow \infty$, $\rho_{1+2}(\lambda)$ depends only on 
$\rho_1(\lambda)$ and $\rho_2(\lambda)$ if $A_1$ and $A_2$ are independent random matrices or saying more precisely if they are free. The freeness is a  concept closely related to the independence. The independence itself is not sufficient. The freeness additionally requires a complete lack of angular correlations. Such correlations may appear even for 
independent matrices if they are generated from a matrix-ensemble which
hides some characteristic angular pattern. An example is just the
ensemble of Wigner-L\'evy matrices whose probability measure is not
rotationally invariant. This measure favors some specific angular
directions. In effect, even if one picks at random two matrices from
this ensemble, they both will prefer some angular directions and thus will have some sort of mutual correlations. One can remove the correlations by a uniform angular randomization of the matrices $A_1$ and $A_2$
\begin{equation}
A_{1 {\scriptscriptstyle\boxplus} 2} = O_1 A_1 O_1^T + O_2 A_2 O_2^T
\label{M12free}
\end{equation}
where $O_i$'s are random orthogonal matrices with a uniform probability 
measure in the group of orthogonal matrices\footnote{For hermitian matrices $A_i$ one uses unitary rotations $A_{1+2} = U_1 A_1 U_1^\dagger + U_2 A_2 U_2^\dagger$}. 
The matrix $O_i A_i O_i^T$ has exactly the same eigenvalue content as $A_i$. Actually to achieve the effect of the angular decorrelation it is sufficient to rotate only one of $A_i$'s. This type of addition is called a free addition and we denote it by
${\scriptstyle \boxplus}$. Of course if $A_i$'s are generated from an ensemble with a rotationally invariant probability measure, as for instance $DA \exp -{\rm tr} V(A)$, then the two types of 
additions (\ref{M12}) and (\ref{M12free}) are identical and $\rho_{1+2}(\lambda)=\rho_{1 {\scriptscriptstyle\boxplus} 2}(\lambda)$. Otherwise they are different
and $\rho_{1+2}(\lambda ) \ne \rho_{1 {\scriptscriptstyle\boxplus} 2}(\lambda)$. 
For example a sum of independent identically distributed Wigner-L\'evy matrices 
(\ref{M12}) is a Wigner-L\'evy matrix which is not rotationally invariant while 
a free sum of Wigner-L\'evy matrices is a rotationally invariant matrix, which has a different 
eigenvalue density, so in this case $\rho_{1+2}(\lambda)\ne \rho_{1 {\scriptscriptstyle\boxplus} 2}(\lambda)$. Generally, for large matrices the eigenvalue density $\rho_{1\boxplus 2}(\lambda)$ 
of a free sum depends only on $\rho_1(\lambda)$ and $\rho_2(\lambda)$ 
and it can be uniquely determined from them \cite{VDN}. 

A theory of free addition was actually developed in probability theory of non-commutative objects (operators) long before random matrices entered the scene \cite{Voi1}. It was originally formulated in terms of von Neumann algebras equipped with a trace-like normal state $\tau$, which was introduced to generalize the concept of uncorrelated variables known from a classical probability.  More specifically in classical probability two real random variables $x_1$ and $x_2$ are uncorrelated if the correlation function, calculated as the expectation value $E(\hat{x}_1 \hat{x}_2) = 0$ for centered variables, $\hat{x}_i = x_i - E(x_i)$, vanishes. In free probability, by analogy, elements $X_i$ are free if  $\tau (\hat{X}_{\pi(1)} \hat{X}_{\pi(2)} \ldots \hat{X}_{\pi(m)} ) = 0$ for any permutation $\pi$ of the corresponding centered elements $\hat{X}_i = X_i - \tau(X_i) \cdot 1$. 
A link to large matrices was discovered later \cite{Voi2} when it was realized that a non-commutative probability space with free elements $X_i$'s can be mapped onto a set of large $N \times N$ matrices, $N\rightarrow \infty$, of the form $X_i  = U_i D_i U_i^\dagger$, where $D_i$ are diagonal real matrices and $U_i$  random unitary (or orthogonal, $O_i$) matrices distributed with a uniform probability measure on the group. In this mapping the state function $\tau$ corresponds to a standard normalized trace operation $\tau(\ldots) \leftrightarrow \frac{1}{N} {\rm tr}(\ldots)$. This observation turned out to be very fruitful both for free probability and for theory of large random matrices since one could successfully apply results of a free probability to random matrices and vice versa \cite{VDN}. Free probability and its relation to random matrices and planar combinatorics for the case when all moments of the  probability distribution exist \cite{Spei,NS} is discussed in detail in Chapter 22. In this section we concentrate on heavy-tailed free distributions for which higher moments do not exist. In particular we shall discuss free stable laws 
and their matrix realizations. Before we do that we recall basic concepts to make the discussion of this section self-contained.

Actually the law of addition of free random variables is in many respects analogous to the law of addition of independent real random variables. We will therefore begin by briefly recalling the law of addition in  classical probability and by analogy describe the corresponding steps in free probability.  
The p.d.f. $p_{1+2}(x)$ of a sum $x=x_1+x_2$ of independent real random
variables $x_1$ and $x_2$ is given by a convolution of the individual 
p.d.f.'s $p_{1+2}(x) = (p_1*p_2)(x)$. Thus, the characteristic function 
$\widehat{p}(k) = \int dk e^{ikx} p(x)$ for the sum is a product of the corresponding characteristic functions $\widehat{p}_{1+2}(k) = \widehat{p}_1(k) \widehat{p}_2(k)$. This is more conveniently expressed in terms of a cumulant-generating function ($c$-transform) defined as $c(k) = \ln \widehat{p}(k)$, as a simple additive law 
\begin{equation}
c_{1+2}(k) = c_1(k) + c_2(k) .
\label{c12}
\end{equation}
It turns out that one can find a corresponding object in free probability \cite{VDN}, a free cumulant-generating function alternatively called $R$-transform, for which the free addition (\ref{M12free}) leads to a corresponding additive rule
\begin{equation}
R_{1 {\scriptscriptstyle\boxplus} 2}(z) = R_1(z) + R_2(z) \ .
\label{R12}
\end{equation}
For a given eigenvalue density $\rho(\lambda)$ one defines a moment-generating 
function as a Cauchy transform of the eigenvalue density \cite{VDN}
\begin{equation}
G(z) = \int_{-\infty}^{+\infty} \frac{\rho(\lambda) d\lambda}{z-\lambda}
\label{G}
\end{equation}
known also as the resolvent or Green function. The free-cumulant-generating function ($R$-transform) is related to the Green function as
\begin{equation}
z = G\left(R(z) + \frac{1}{z}\right) \ .
\label{RG}
\end{equation}
The last equation can be inverted for $G(z)$
\begin{equation}
z = R(G(z)) + \frac{1}{G(z)} \ .
\label{GR}
\end{equation}
In short $z \rightarrow R(z) + 1/z$ is the inverse function of $z \rightarrow G(z)$ and thus for a given resolvent one can determine the $R$-transform 
and vice versa. Using the addition law (\ref{R12}) one can now give a step-by-step algorithm to calculate the eigenvalue density $\rho_{1 {\scriptscriptstyle\boxplus} 2}(\lambda)$ of the free sum (\ref{M12free}) from $\rho_1(\lambda)$ and $\rho_2(\lambda)$. First one determines the resolvents $G_1(z)$ and $G_2(z)$ using (\ref{G}), then the corresponding $R$-transforms $R_1(z)$ and $R_2(z)$ using (\ref{RG}) and finally $R_{1 {\scriptscriptstyle\boxplus} 2}(z)$ using the addition law (\ref{R12}). Having found $R_{1 {\scriptscriptstyle\boxplus} 2}(z)$ one proceeds in the opposite order. One reconstructs the corresponding resolvent $G_{1 {\scriptscriptstyle\boxplus} 2}(z)$ (\ref{GR}) and then the density 
$\rho_{1  {\scriptscriptstyle\boxplus} 2}(\lambda)$ by the inverse of (\ref{G})
\begin{equation}
\rho(\lambda) = - \frac{1}{\pi}  {\rm Im} G(\lambda + i  0^+)
\label{rhoG}
\end{equation}
which follows from the relation $(x+0^+)^{-1} = P.V. x^{-1}  - i\pi \delta(x)$. This completes the task of calculating the eigenvalue density 
of a free sum $\rho_{1  {\scriptscriptstyle\boxplus} 2}(\lambda)$ from $\rho_1(\lambda)$ and $\rho_2(\lambda)$. This procedure can be fully automatized and it actually has been implemented for a certain class of matrices \cite{Rao}.

The correspondence between the laws of addition (\ref{c12}) and (\ref{R12}) and between the logical structures behind these laws in  classical and free probability has also profound theoretical implications. One of them is a bijection between infinitely divisible laws of classical probability 
and the laws in free probability \cite{BV}. Using this bijection one can derive the $R$-transform for stable laws in free probability\footnote{We give only the standardized version which corresponds to 
the unit range. An $R$-transform with a range $r$ can be obtained from the standardized one by a rescaling $R_r(z) = r R(r z)$.} \cite{BPB}
\begin{equation}
R(z) = \left\{\begin{array}{ll}  b z^{\alpha-1} & {\rm for} \ \alpha \in (0,2] \ {\rm and} \ \alpha \ne 1 \\
-i(1\!+\!\beta) - (2\beta /\pi) \ln z & {\rm for}  \ \alpha=1 \\
\end{array} \right. 
\label{Rstable}
\end{equation}
where
$b = -e^{i \alpha (1 + \beta) \pi/2}$ for $\alpha \in (0,1)$ and $b=e^{i(\alpha - 2)(1+\beta) \pi/2}$ for $\alpha \in (1,2]$.
The stable $R$-transforms (\ref{Rstable}) are in one-to-one correspondence with the $c$-transforms of L\'evy distributions (\ref{charL}) and they fully classify all free stable laws and allow one to determine the free probability densities for these stable laws. These densities are equal to the eigenvalue densities of free L\'evy matrices being matrix realizations of the stable free random variables.

Let us illustrate how this procedure works for a stable law with the stability index $\alpha=2$ and unit range. Using (\ref{Rstable}) we have $R(z) = z$. Inserting this to (\ref{GR}) we obtain an equation for the resolvent 
$G(z) = z - 1/G(z)$ which gives in the upper complex half-plane 
$G(z)= (z - \sqrt{z^2 - 4})/2$. Finally using (\ref{rhoG}) we find a Wigner law
$\rho(\lambda) = \sqrt{4-\lambda^2}/2\pi$ with $\sigma=1$. We see that the Wigner law is equivalent in free probability to the normal law in classical probability. Secondly consider the case for $\alpha=1$, $\beta=0$. Proceeding in the same way as above we find respectively
\begin{equation}
R(z) = -i \ , \quad G(z) = \frac{1}{z + i} \ , \quad
\rho(\lambda) = \frac{1}{\pi} \frac{1}{1+\lambda^2} \ .
\end{equation}
This case is special because the stable density is identical in classical and free probability. For other values of $\alpha$ one can find the free stable laws by applying the equation (\ref{GR}) to the $R$-transform (\ref{Rstable}) which gives the following equation for the resolvent 
\begin{equation}
b G^\alpha(z) - z G(z) + 1 = 0 \ .
\label{Gfree}
\end{equation} 
This equation can be solved analytically for a couple of values of the
parameter $\alpha$ for which it is just a quadratic, cubic or quartic equation.
The solution can be then used to calculate the eigenvalue density (\ref{rhoG}).
For other values the eigenvalue density can be determined numerically.

The equation (\ref{Gfree}) may be used to extract the asymptotic behavior of 
the corresponding eigenvalue density (\ref{rhoG}). We will give the result only for the symmetric case $(\beta=0)$ and for the range $R=1$. For small eigenvalues $|\lambda| \rightarrow 0$ it reads \cite{BJ2NPZ} 
\begin{equation}
\rho(\lambda) = \frac{1}{\pi} \left( 1 - \frac{3-\alpha}{2\alpha^2} \lambda^2 + \ldots \right)
\end{equation}
while for large ones, $|\lambda| \rightarrow \infty$
\begin{equation}
\rho(\lambda) \sim \frac{1}{\pi} \sin\left(\frac{\alpha \pi}{2}\right) |\lambda|^{-\alpha-1} \ .
\label{large_lambda_free}
\end{equation}
The distribution has a smooth quadratic maximum at $\lambda=0$ while for
large $\lambda$ it has heavy power-law tails with the same power  as the
corresponding stable law for real variables. For $\alpha=2$ the tails disappear.

\begin{figure}
\begin{center}
\epsfig{figure=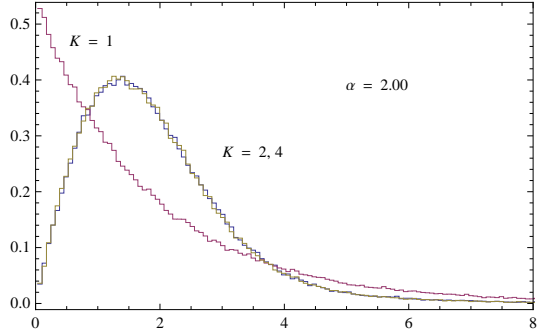,width=7cm}
\end{center}
\caption{\label{repulsion} Level spacing histograms of the matrices
$A_{{\scriptscriptstyle\boxplus} K}$ (\ref{odo_k}) of size $200\times 200$ 
obtained from diagonal matrices 
$D_i$ whose  eigenvalues were generated independently from a Wigner
semicircle. For $K=1$ the histogram properly 
reflects the Poissonian
nature of eigenvalues of a single matrix $D_i$. 
For $K>1$ (already for $K=2$) the histogram
has a shape of the Wigner surmise characteristic
for a typical eigenvalue repulsion.} 
\end{figure}

The procedure described above to derive the free probability density $\rho(\lambda)$ of free stable laws is solely based on relations between the $R$-transform, the resolvent and the density. It does not involve any matrix calculations. Free random matrices appear as a representation of these free random variables and 
correspond to infinitely large random matrices with a given eigenvalue density $\rho(\lambda)$ and with a rotationally invariant probability measure.
Actually there are many different matrix realizations of the same free random variable. We shall below discuss two simplest ones which have completely different microscopic properties. The most natural realization is a matrix
generated by the uniform angular randomization $ODO^T$ of a large diagonal matrix $D$ of size $N \rightarrow\infty$ which has i.i.d. random variables on the diagonal. If we choose the p.d.f. of the diagonal elements as $\rho(\lambda)$, which corresponds to a free stable law, we obtain a matrix realization of a free random variable which is stable under addition. Clearly, the eigenvalues of this matrix are by construction uncorrelated. We can now independently generate many such matrices $O_i D_i O_i^T$, $i=1,\ldots, K$. 
Due to the stability also the following sum
\begin{equation}
A_{ {\scriptscriptstyle\boxplus} K} = \frac{1}{K^{1/\alpha}} \sum_{i=1}^K O_i D_i O_i^T
\label{odo_k}
\end{equation}
has exactly the same eigenvalue density $\rho(\lambda)$ as each term in the sum, so it is a representation of the same free random variable. However the sum $A_{ {\scriptscriptstyle\boxplus} K}$ belongs to a different microscopic class since its eigenvalues repel each other contrary to eigenvalues of each matrix in the sum which are uncorrelated by construction. Already for $K=2$ one observes a standard repulsion characteristic for invariant ensembles as illustrated in Figure \ref{repulsion}.
It is tempting to conjecture that for $K\rightarrow \infty$ the sum $A_{ {\scriptscriptstyle\boxplus} K}$  
becomes  a random matrix which maximizes entropy for the stable eigenvalue density $\rho(\lambda)$. The probability measure for
such matrices is known to be rotationally invariant $d \mu(A)=DA \exp -N{\rm tr} V(A)$ and to have a potential $V(\lambda)$ which is related to $\rho(\lambda)$ as \cite{Balian}
\begin{equation}
V'(\lambda) = 2 P.V. \int d\zeta \frac{\rho(\zeta)}{\lambda - \zeta} = 
G(\lambda+i0^+)+G(\lambda-i0^-) \ .
\end{equation}
 For free stable laws the resolvent $G(\lambda)$ 
is given by a solution of (\ref{Gfree}). In particular, for $\alpha=2$ the
last equation gives a quadratic potential $V(\lambda) = \lambda^2/2$, while for $\alpha=1$
and $\beta=0$ a logarithmic one $V(\lambda) = \ln(\lambda^2 + 1)$. 
The potential can also be determined for other values of $\alpha$. Generally, for large $|\lambda| \rightarrow \infty$ it behaves as
\begin{equation}
V(\lambda) = 2 \ln \lambda - 2 \alpha^{-1} {\rm Re} \, \left( b \lambda^{-\alpha} \right) 
+ \ldots 
\end{equation}  
For maximal entropy random matrices one can also find the joint
probability which takes a standard form
\begin{equation}
\rho(\lambda_1,\ldots,\lambda_N) = C e^{-N \sum_i V(\lambda_i)} 
\prod_{i < j} (\lambda_i - \lambda_j)^\beta
\label{joint}
\end{equation}
where $C$ is a normalization and $\beta$ as usual is equal $1$ or $2$
for orthogonal or unitary invariant ensemble respectively. For free stable laws the potential $V(\lambda)$ assumes however a highly non-standard
 non-polynomial form \cite{BJ2NPZ}.

\begin{figure}
\begin{center}
\epsfig{figure=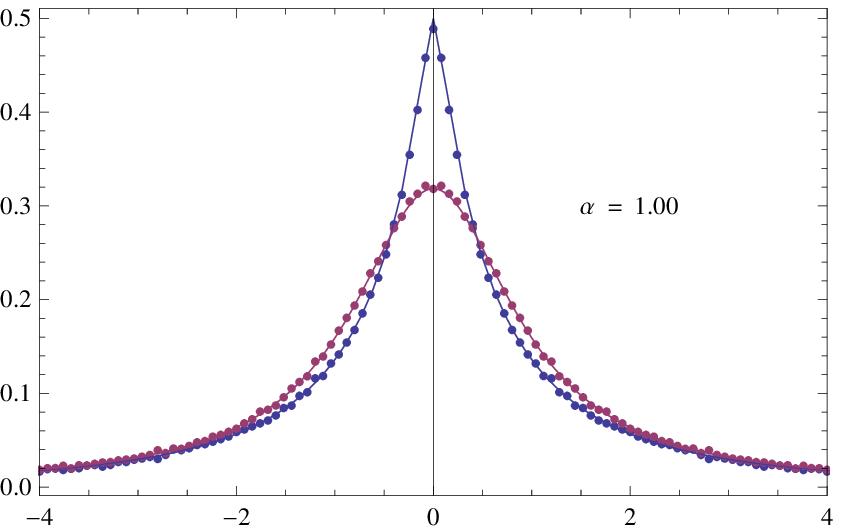,width=6.4cm} 
\epsfig{figure=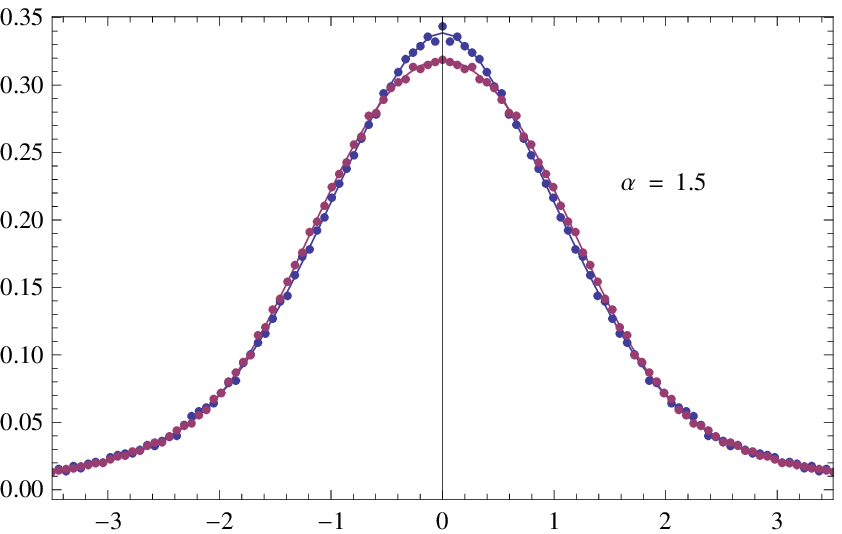,width=6.4cm}
\end{center}
\caption{\label{wigner_and_free}
Eigenvalue density for Wigner-L\'evy matrices
and corresponding free L\'evy matrices for $\alpha=1.00$ (left) and $\alpha=1.50$ (right) 
are represented by solid line. The corresponding Monte-Carlo histogram for a 
free sum (\ref{fsummat}) of $K=32$ Wigner-L\'evy matrices of size
$N=200$ is  shown. For comparison we chose the range of the
Wigner-L\'evy matrix to be $R=(\Gamma(1+\alpha))^{-1/\alpha}$ since then
the asymptotic behavior of the eigenvalue density for Wigner-L\'evy
matrices  (\ref{large_lambda_WL}) is identical as for the standardized free random variables (\ref{large_lambda_free}).}
\end{figure}

The stable laws in a free probability play a similar role as the corresponding stable laws in classical probability where a sum of i.i.d. centered random variables $s_n=(x_1+\ldots+x_n)/n^{1/\alpha}$ is known to become an $\alpha$-stable random variable for $n\rightarrow \infty$. We expect a similar effect for random matrices \cite{HP}. For example, if one generates a sequence of $K$ independent Wigner-L\'evy matrices $(A_1,A_2,\ldots,A_K)$ and a sequence independent random orthogonal matrices $(O_1,O_2,\ldots,O_K)$ and one 
forms a sum
\begin{equation}
A_{ {\scriptscriptstyle\boxplus} K} = \frac{1}{K^{1/\alpha}} \sum_{i=1}^K O_i A_i O_i^T 
\label{fsummat}
\end{equation}
in analogy to (\ref{odo_k}) one expects that for large $K$ the sum will become a free random matrix with an eigenvalue density given by a free $\alpha$-stable law \cite{BJNPZ}. Actually in practice the convergence to the limiting distribution is very fast. We observe that already for $K$ or order $10$ and $N$ of order $100$ the eigenvalue density of the matrix $A_{{\scriptscriptstyle\boxplus} K}$ does not significantly differ from the stable density $\rho(\lambda)$ for free random variables (see Figure \ref{wigner_and_free}). 
If we skipped random rotations in (\ref{fsummat}) and just added many Wigner-L\'evy matrices $(A_1+\ldots+A_K)/K^{1/\alpha}$ we would obtain a Wigner-L\'evy matrix. As we mentioned
already, the independence of the matrices $A_i$ itself is not sufficient to make the addition free and only the rotational randomization brings the sum to the universality class of free random variables. The comparison of the eigenvalue density of the Wigner-L\'evy and the corresponding free L\'evy matrices is shown in Figure \ref{wigner_and_free}.

\sect{Heavy tailed deformations}

In this section we discuss ensembles of random matrices obtained from the
standard ensembles by a reweighting of the probability measure. We will begin
by sketching the idea for real random variables where the procedure is simple 
and well known and then we will generalize it to random matrices. 
We will in particular concentrate on heavy-tailed deformations of the 
probability measure of the Wishart ensemble which are applicable to the 
statistical multivariate analysis of heavy tailed data. 

Consider a random variable $x$ constructed as a product 
$x= \sigma\xi$ where $\xi$ is a normally distributed 
variable ${\cal N}(0,1)$ and $\sigma \in (0,\infty)$ is
an independent random variable representing a fluctuating 
scale factor having a p.d.f. $f(\sigma)$. One can think of $x$ as a Gaussian
variable ${\cal N}(0,\sigma^2)$ for which the variance itself is a random variable. Obviously the p.d.f. of $x$ can be calculated as
\begin{equation}
p(x) =  \int_0^\infty d\sigma f(\sigma) \frac{1}{\sqrt{2\pi}\sigma}
e^{-\frac{x^2}{2 \sigma^2}} \ .
\label{sf}
\end{equation}
Choosing appropriately the frequency function $f(\sigma)$ one can thus
model the p.d.f. of $x$. For example for \cite{TVT,BBP,Abul}
\begin{equation}
f(\sigma) = \frac{1}{\sigma} \frac{2}{\Gamma\left(\frac{\alpha}{2}\right)} 
\left(\frac{a^2}{2\sigma^2}\right)^{\frac{\alpha}{2}} 
e^{-\frac{a^2}{2 \sigma^2}}
\label{fis}
\end{equation}
where $a$ is a constant, the integral (\ref{sf}) takes the following form 
\begin{equation}
p(x) =  \frac{1}{a\sqrt{\pi}}  \frac{1}{\Gamma\left(\frac{\alpha}{2}\right)} 
\int_{0}^{\infty} d \zeta \ \zeta^{\frac{\alpha-1}{2}} e^{-\zeta} 
e^{-\zeta \left(\frac{x}{a}\right)^2} 
= \frac{1}{a\sqrt{\pi}} \frac{\Gamma\left(\frac{\alpha+1}{2}\right)}{\Gamma\left(\frac{\alpha}{2}\right)} \left(1+\left(\frac{x}{a}\right)^2\right)^{- \frac{\alpha+1}{2}}  \ .
\label{student}
\end{equation}
In doing the integral we  changed the integration variable\footnote{The variable $\zeta$ 
has a $\chi^2$ distribution.} to $\zeta=a^2/(2\sigma^2)$. 
For large $|x|$ the p.d.f. has a power-law tail 
$p(x) \sim |x|^{-\alpha-1}$. The variance of the distribution exists only for $\alpha>2$ and is equal to $a^2/(\alpha-2)$. 

If one applies this procedure to each matrix element independently 
$A_{ij} = \sigma_{ij} \xi_{ij}$, $i\le j$, one obtains a Wigner matrix
discussed in the first section of this chapter. One can however construct 
a slightly different matrix $A_N(\sigma)$ whose elements
have the same common random scale factor $A_{ij} = \sigma \xi_{ij}$ and 
differ only by $\xi_{ij}$ which are i.i.d. Gaussian 
random variables ${\cal N}(0,1)$. In the limit $N\rightarrow \infty$ the eigenvalue density of the matrix 
$A_N(\sigma)/\sqrt{N}$ converges to the Wigner semicircle law 
$\rho_\sigma(\lambda) = \sqrt{4\sigma^2 -\lambda^2}/(2\pi\sigma^2)$. The scale factor $\sigma$ changes however from matrix to matrix with the frequency $f(\sigma)$ so in analogy to (\ref{sf}) the effective eigenvalue density in the ensemble of matrices is given by the average of 
the semicircle law over $\sigma$  \cite{BCP1}
\begin{equation}
\rho(\lambda) = \int d\sigma f(\sigma) \rho_\sigma(\lambda) =  
\int_{\frac{\lambda}{2}}^\infty d\sigma f(\sigma) \frac{1}{2\pi \sigma^2}
\sqrt{4\sigma^2 -\lambda^2} \ .
\end{equation}
In particular, for the frequency function (\ref{fis}) we obtain the following eigenvalue density \cite{BBP}
\begin{equation}
\rho(\lambda) = \frac{1}{a} \frac{\sqrt{2}}{\pi \Gamma\left(\frac{\alpha}{2}\right)} 
\int_{0}^{\frac{2a^2}{\lambda^2}} d\zeta 
\ \zeta^{\frac{\alpha-1}{2}} e^{-\zeta} 
\sqrt{1 -\frac{\zeta \lambda^2}{2a^2}} 
\end{equation}
which has power-law tails, $\rho(\lambda) \sim |\lambda|^{-\alpha-1}$, with the same power as the p.d.f. (\ref{student}) for matrix elements. One can generalize the result to other frequency functions \cite{MK}.

Actually exactly the same strategy can be applied to calculate the joint probability since fluctuations of matrix elements are modified by a common scale factor $\rho(\lambda_1,\ldots,\lambda_N) = \int d\sigma f(\sigma) \rho_\sigma(\lambda_1,\ldots,\lambda_N)$, where $\rho_\sigma$ is given by (\ref{joint}) with $V(\lambda) = \lambda^2/(2\sigma^2)$. From the joint probability one can then derive microscopic properties of matrices, including the microscopic correlation functions and the distribution of the largest eigenvalue \cite{BCP2}. 

One can use this procedure to deform probability measures of 
other matrix ensembles as well. In what follows we concentrate on deformations of the Wishart ensemble. The probability measure for a standardized Wishart ensemble of real matrices\footnote{For complex matrices the corresponding measure reads
$d\mu_*(\xi)= \pi^{-NT} e^{-{\rm tr} \xi \xi^\dagger} D\xi $ where 
$D\xi = \prod^{NT}_{it} d {\rm Re} \xi_{it} \; d {\rm Im} {\xi}_{it}$ } 
is given by 
\begin{equation}
d \mu_*(\xi) =  \left(2\pi\right)^{-\frac{NT}{2}} 
e^{-\frac{1}{2} {\rm tr} \ \xi \xi^T} D\xi
\label{stand_wish}
\end{equation}
where $\xi$ is a rectangular matrix $\xi_{it}$, $i=1,\ldots, N$, $t=1,\ldots,T$ and $D\xi = \prod_{i,t}^{N,T} d\xi_{it}$.  The eigenvalue density of the matrix $(1/T) \xi \xi^T$ is known to converge  to the Mar\v{c}enko-Pastur law $\rho_*(\lambda) = \sqrt{(\lambda_+ - \lambda)(\lambda-\lambda_-)}/(2\pi r \lambda)$, where $\lambda_\pm = (1\pm r)^2$ and $r=N/T$ in the limit $N \rightarrow \infty$, $r={\rm const}$
\cite{MP}. The elements of the matrix $\xi$ represent normally distributed fluctuations of uncorrelated random numbers with unit variance. Using now the reweighting method we can consider a matrix $A_{it} = \sigma \xi_{it}$ with a common fluctuating scale factor being an independent random variable with a p.d.f. $f(\sigma)$. 
The effective probability measure can be easily derived from (\ref{stand_wish}) and reads
\begin{equation}
d \mu(A) = DA \int d\sigma f(\sigma) \sigma^{-NT}
e^{-\frac{1}{2\sigma^2} {\rm tr} A A^T}
\label{deform_wish}
\end{equation}
where the factor $\sigma^{-NT}$ comes from the change of variables in the 
measure $DA = \sigma^{NT} D\xi$. In particular for the frequency function 
(\ref{fis}) this gives the measure of a multivariate Student's distribution\footnote{An analogous expression \cite{TVT,BBP} can be obtained for 
GOE and GUE Wigner matrices reweighted with the frequency function (\ref{fis}).}
\begin{equation}
d \mu(A) = DA  \frac{\Gamma\left(\frac{\alpha+NT}{2}\right)}{
(a \sqrt{\pi})^{NT}\Gamma\left(\frac{\alpha}{2}\right)} \left(1+\frac{{\rm tr} \; AA^T}{a^2}\right)^{- \frac{\alpha+NT}{2}} \ .
\label{mstudent}
\end{equation}
The eigenvalue density of the matrix $(1/T) XX^T$, where $X$ is generated
with the probability measure given above can be obtained by the same 
reweighting method as before. The eigenvalue density 
of the Gaussian part in (\ref{deform_wish}) is 
$\rho_\sigma(\lambda) = \rho_*(\lambda/\sigma^2)/\sigma^2 = 
\sqrt{(\sigma^2\lambda_+ - \lambda)(\lambda-\sigma^2\lambda_-)}/(2\pi r \sigma^2 \lambda)$. It has to be reweighted with the frequency function (\ref{fis}) 
$\rho(\lambda) = \int d\sigma f(\sigma) \rho_\sigma(\lambda)$.  
The result reads \cite{BGW}
\begin{equation}
\rho(\lambda) = \frac{\left(\frac{\alpha}{2}\right)^{\alpha/2}}{2\pi r \Gamma\left(\frac{\alpha}{2}\right)}  \lambda^{-\alpha/2 - 1} 
\int^{\lambda_+}_{\lambda_-} \sqrt{(\lambda_+ - \zeta)(\zeta-\lambda_-)} \
e^{-\frac{\alpha\zeta}{2\lambda}} \zeta^{\alpha/2-1} d\zeta  \ .
\end{equation}
The support of this eigenvalue distribution is infinite. The exponent $\alpha/2$ 
of the tail is a half of the exponent $\alpha$ of p.d.f. for the matrix elements $X$ as one can expect for a matrix $(1/T) XX^T$ which is a ``square'' of $X$. The reweighting can be applied to derive the corresponding joint probability function and to determine the microscopic correlations 
of the deformed Wishart ensembles \cite{AV}.

This result can be generalized in a couple of ways. One can change
the frequency 
function $f(\sigma)$ \cite{AAV} but one can also change the relation
between the 
$A$ and $\xi$ matrix from $A_{it} = \sigma \xi_{it}$ to, for instance, 
$A_{it} = \sigma \sum_j S_i O_{ij} \xi_{jt}$ 
where $O_{ij}$ is a rotation matrix and $S_i$ is a vector of positive numbers.
The matrix $O$ and the vector $S$ are fixed in this construction and
the only fluctuating elements are $\xi_{it}$ which are i.i.d. 
${\cal N}(0,1)$ and $\sigma$ which is an independent random variable
with a given p.d.f. $f(\sigma)$, as before. The interpretation of the
construction is clear. The factors $S_i$'s change the
scale of fluctuations and the matrix $O$ rotates the main axes.
If one applies it to (\ref{stand_wish}) one will obtain a deformed measure 
(\ref{mstudent}) where ${\rm tr A A^T}$ will be substituted by ${\rm tr A C^{-1} A^T}$. The matrix $C_{ij} = \sum_k O_{ik} S_k^2 O_{kj}$ introduces 
explicit correlations between the degrees of 
freedom. In a similar way one can introduce correlations $C_{tt'}$ between $A_{it}$ and $A_{it'}$ at different times $t$ and $t'$ \cite{BGW}. 

Another interesting generalization of the reweighting procedure is to consider 
a matrix $A_{it} = \sigma_i \xi_{it}$ where now the scale factors
$\sigma_i$ are independent random variables for each row \cite{BBP2}. This case 
corresponds to a Wishart ensemble with a fluctuating covariance matrix 
$C_{ij} = \delta_{ij} \sigma_i^2$ where $\sigma_i$ are i.i.d. random variables with a given p.d.f. $f(\sigma)$. This problem can be solved 
analytically thanks to an explicit relation between the  Greens functions and eigenvalue densities of the matrices $C$ and $(1/T) XX^T$ \cite{BGJJ}.

The idea of reweighting is quite general. So far we have described
reweighting through the scale parameter $\sigma$ but one can use other
quantities as a basis for the reweighting scheme as well. For example
one can use the trace $t={\rm tr} XX^T$ of the whole matrix. In this scheme the idea is to calculate quantities for the ensemble with a probability measure\footnote{Alternatively one can use 
$d\mu'_t(X) = \theta ({\rm tr} XX^T -t) DX$, where $\theta(x)$ is the Heaviside step function \cite{BBP}.}
$d\mu_t(X) = DX \delta(t - {\rm tr} XX^T)$ and then to reweight them using a frequency function $g(t)$ to obtain the corresponding values for the ensemble with the measure $d\mu(X) = DX g({\rm tr} XX^T)$ \cite{ACMV,Abul}. It turns out that the first step, that is the calculations for the fixed trace ensemble, can be done analytically so also this procedure gives a practical recipe to handle ``non-standard'' ensembles.  Of course it works only for  ensembles for which the measure depends on ${\rm tr} XX^T$ as for instance (\ref{mstudent}). In particular it can be applied to the multivariate Wishart-Student ensembles \cite{BGW}.

\sect{Summary}
Heavy-tailed random matrices is a relatively new branch of random matrix theory. In this chapter we discussed several matrix models belonging to this class and presented methods of integrating them. We believe that the models, methods and concepts can be applied to many statistical problems where non-Gaussian effects play an important role.

\ \\
{\sc Acknowledgements} 

\noindent
We thank G. Akemann, J.P. Bouchaud, A. G\"orlich, R.A. Janik, A. Jarosz, M.A. Nowak, G. Papp, P. Vivo, B. Waclaw and I. Zahed for many interesting discussions. This work was supported by the Marie Curie ToK project ``COCOS'', No.~MTKD-CT-2004-517186, the EC-RTN Network ``ENRAGE'', 
No.~MRTN-CT-2004-005616 and the Polish Ministry of Science Grant
No.~N~N202~229137 (2009-2012).


\begin{thebibliography}{Abc84a}
\bibitem[Abu05]{Abul} A. Y. Abul-Magd, Phys. Rev. E {\bf 71} (2005) 066207. 
\bibitem[Abu09]{AAV} A.Y. Abul-Magd, G. Akemann, P. Vivo,  J. Phys. A: Math. Theor. {\bf 42} (2009) 175207.
\bibitem[Ake99]{ACMV} G. Akemann, G.M. Cicuta, L. Molinari, G. Vernizzi, Phys. Rev. E {\bf 59} (1999) 1489; Phys. Rev. E {\bf 60} (1999) 5287.
\bibitem[Ake08]{AV} G. Akemann and P. Vivo, J. Stat. Mech. (2008) P09002.
\bibitem[Bal68]{Balian} R. Balian, Nuovo Cimento B {\bf 57} (1968) 183. 
\bibitem[Ber93]{BV} H. Bercovici and D. Voiculescu, Ind. Univ. Math. J. {\bf 42} 
(1993) 733. 
\bibitem[Ber99]{BPB} H. Bercovici and V. Pata, Annals of Mathematics {\bf 149} 
(1999) 1023; Appendix by P. Biane. 
\bibitem[Ber04]{BBP}  A.C. Bertuola, O. Bohigas, and M.P. Pato, Phys. Rev E {\bf 70} (2004) 065102(R).
\bibitem[Bir07a]{BBP1} G. Biroli, J.-P. Bouchaud, and M. Potters, Europhys. Lett. {\bf 78} (2007) 10001. 
\bibitem[Bir07b]{BBP2} G. Biroli, J.-P. Bouchaud, and M. Potters, Acta Phys. Pol. B {\bf 38} (2007) 4009.
\bibitem[Boh08]{BCP1} O. Bohigas, J.X. de Carvalho and M.P. Pato, Phys. Rev. E {\bf 77} (2008) 011122.
\bibitem[Boh09]{BCP2} O. Bohigas, J.X. de Carvalho and M.P. Pato, 
Phys. Rev. E {\bf 79} (2009) 031117.
\bibitem[Bou97]{BM} J.-P. Bouchaud and M. M\'ezard, J. Phys. A, Math. Gen. 
{\bf 30} (1997) 7997. 
\bibitem[Bur02]{BJ2NPZ} Z. Burda, R.A. Janik, J. Jurkiewicz, M.A. Nowak, G. Papp and I. Zahed, Phys. Rev. E {\bf 65} (2002) 021106.
\bibitem[Bur04]{BGJJ} Z. Burda, A. G\"{o}rlich, A. Jarosz, J. Jurkiewicz, Physica A {\bf 343} (2004) 295.
\bibitem[Bur06]{BGW} Z. Burda, A. G\"{o}rlich, B. Waclaw, Phys. Rev. E {\bf 74} (2006) 041129.
\bibitem[Bur07]{BJNPZ} Z. Burda, J. Jurkiewicz, M. A. Nowak, G. Papp and I. Zahed, Phys. Rev. E {\bf 75} (2007) 051126;  arXiv:cond-mat/0602087.
\bibitem[Ciz94]{CB} P. Cizeau and J.P. Bouchaud, Phys. Rev. E {\bf 50} (1994) 1810. 
\bibitem[Fel71]{Feller} W. Feller, {\it An Introduction to Probability Theory and Its Applications}, Wiley, 3rd Edition, New York 1971.
\bibitem[Gne68]{GK} B.V. Gnedenko, A. N. Kolmogorov, {\em Limit distributions for sums of independent random variables}, Revised Edition
Addison-Wesley,  Cambridge 1968.
\bibitem[Hia00]{HP} F. Hiai and D. Petz, {\em The Semicircle Law, Free Random Variables and Entropy}, Am. Math. Soc., Providence 1992.
\bibitem[Mar67]{MP} V.A. Mar\v{c}enko and L. A. Pastur, Math. USSR-Sb, {\bf 1}, (1967) 457.
\bibitem[Meh04]{Mehta} M.L. Mehta, {\it Random Matrices}, Academic Press, 3rd Edition, London 2004. 
\bibitem[Mut05]{MK} K.A. Muttalib and J.R. Klauder, Phys. Rev. E {\bf 71}, (2005) 055101(R).
\bibitem[Nic06]{NS} A. Nica and R. Speicher, {\em Lectures on the Combinatorics of Free Probability}, London Mathematical Society Lecture Note Series, vol. 335, 
Cambridge University Press, 2006.
\bibitem[Nol10]{Nolan} J.P. Nolan, {\it Stable Distributions - Models for Heavy Tailed Data}, Birkh\"auser, Boston 2010; http://academic2.american.edu/$\sim$jpnolan.
\bibitem[Pas72]{Pastur} L.A. Pastur: Teor. Mat. Fiz.,  {\bf 10} (1972) 102. 
\bibitem[Rao06]{Rao} N.R. Rao, {\em RMTool: A random matrix and free probability calculator in MAT-LAB}; http://www.mit.edu/~raj/rmtool/. 
\bibitem[Sos04]{Sosh} A. Soshnikov, Elect. Comm. in Probab. {\bf 9} (2004) 82.
\bibitem[Spe94]{Spei} R. Speicher, Math. Ann. {\bf 298} (1994) 611.
\bibitem[Tos04]{TVT} F. Toscano, R.O. Vallejos and C. Tsallis, Phys. Rev. E {\bf 69} (2004) 066131.
\bibitem[Tra94]{TW1} C.A. Tracy and H. Widom, Commun. Math. Phys. {\bf 159} (1994) 151. 
\bibitem[Tra96]{TW2} C.A. Tracy and H. Widom, Commun. Math. Phys. {\bf 177} (1996) 724. 
\bibitem[Voi85]{Voi1} D.V. Voiculescu, in {\em Operator algebras and their connections with topology and ergodic theory}, (Busteni, 1983), Lecture Notes in Math. Series, vol. {\bf 1132}, 556, Springer, New York 1985. 
\bibitem[Voi91]{Voi2} D.V. Voiculescu, Invent. Math. {\bf 104} (1991), 201220. 
\bibitem[Voi92]{VDN} D.V. Voiculescu, K.J. Dykema and A. Nica, {\it Free Random 
Variables}, Am. Math. Soc., Providence 1992.
\end{thebibliography}
\end{document}